# The Extended Diode Equation

Jianhong Yang*, Gonghao Sun, and Tinglan Li

*School of Physical Science and Technology, Lanzhou University, Lanzhou 730000, China*

*Corresponding author: yangjh@lzu.edu.cn

***Abstract***—For more than 75 years, the p–n junction current-voltage characteristic has been derived by treating majority carriers as an inert background and conflating the junction voltage with the applied bias. Here we show that the base electric field, generated by the coupling of excess majority and minority carriers, fundamentally alters the internal potential distribution. This self-consistent framework yields a closed-form extended diode equation that reproduces the high-injection roll-off without invoking an external series resistance, thereby eliminating an empirical patch that has persisted in the classical theory.

***Introduction***—The p–n junction is the foundational element of semiconductor technology, and the theory formulated by Shockley in 1949 [1,2] remains the standard account of its current–voltage characteristic in every textbook [3,4,18,19]. The theory derives the ideal diode law from minority-carrier diffusion alone, resting on three simplifying premises: the majority carriers remain an inert background clamped at their equilibrium concentration, the quasi-neutral base carries no electric field, and the junction voltage $V_j$ equals the externally applied bias $V_a$. These premises are mutually consistent only at low injection. Under the high-bias, high-injection conditions in which power diodes, solar cells, and junction-based sensors actually operate, the theory fails for all of them at once; its most visible defect, the high-injection current roll-off, is repaired only by appending an empirical series resistance that lies outside the physics of the junction [3–5,18,19].

The roll-off itself, the drift of the ideality factor from $\mathrm{n} = 1$ toward $\mathrm{n} = 2$, has been documented for decades in power rectifiers and p–i–n structures [6–9,20], and case-specific analyses exist for particular high-current regimes [8–10,20,21]. What has been missing is not a description of any single regime, but a unified junction law: one that identifies which assumption of the classical theory breaks, repairs it from first principles, contains the Shockley equation as a demonstrable limit, and makes quantitative, falsifiable predictions beyond it. In this Letter we provide such a law.

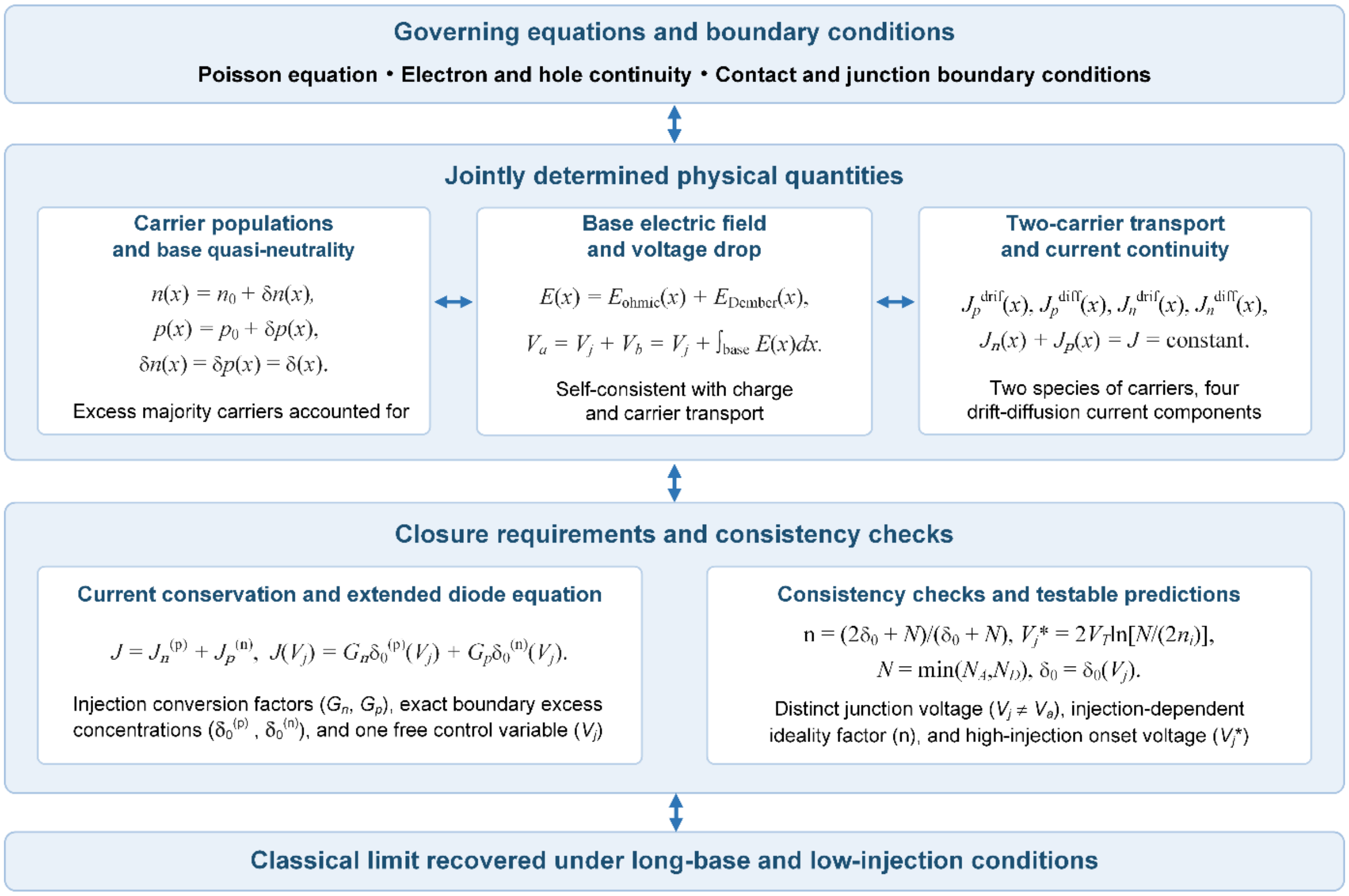

**FIG. 1.** Governing equations (top) jointly determine three groups of internal quantities (second row); closure by current conservation yields the extended diode equation, while consistency checks yield three parameter-free relations (third row); the classical theory is recovered as the stated limit (bottom).

The full derivation, together with the device-level consequences for power dissipation and base transit time, is given in a companion article [22]. Its starting point is a population that the classical theory discards: the excess majority carriers. We show that they are not optional corrections but electrostatically forced transport participants; incorporating them without introducing any assumption beyond the classical ones yields a self-consistent two-carrier drift–diffusion framework (Fig. 1) whose device-level closure is a closed-form extended diode equation valid across forward and reverse bias, low and high injection, and arbitrary base length.

***Excess majority carriers and the carrier profile.***—Injecting minority carriers into a quasi-neutral base without an equal excess majority population would create a space charge of the order of the doping density, which Poisson's equation forbids: the electrostatics that defines a quasi-neutral region forces an equal excess population of the majority species to accompany the injected one [11,12]. Full numerical solution of the coupled drift–diffusion–Poisson system confirms that the relative imbalance $(\delta p - \delta n)/\delta p$ remains of the order of the dielectric-to-lifetime ratio $\tau_d/\tau_p \sim$ $10^{-7}$ for Si under all bias polarities and injection levels [13]. We therefore impose, not as an approximation but as a first-principles bound,

$$\delta p(x) = \delta n(x) \equiv \delta(x). \tag{1}$$

With the two species sharing one profile, the electron and hole continuity equations combine into the ambipolar diffusion equation

$$\frac{d^2\delta}{dx^2} = \frac{\delta}{L_a^2}, \qquad L_a = \sqrt{D_a \tau_a}, \tag{2}$$

where $D_a$ and $\tau_a$ are the ambipolar diffusivity and lifetime [11,12]. Two points deserve emphasis. First, $D_a$ is not the minority-carrier diffusivity: the internal field that maintains quasi-neutrality speeds the slower species and retards the faster one, and this electrostatic coupling is folded into $D_a$. Second, eliminating the net-drift term of the ambipolar *transport* equation is used only to obtain the profile; the physically real drift currents of both species are retained in full below, exactly where the classical treatment discards them. Taking the n-type base of a uniformly doped abrupt junction (the p base is fully symmetric), with $x = 0$ at the space-charge-region (SCR) edge and $x = W$ at the ohmic contact, the solution is the unified hyperbolic profile

$$\delta(x) = \delta_0 \frac{\sinh[(W - x)/L_a]}{\sinh(W/L_a)}, \tag{3}$$

which reduces to the long-base exponential for $W \gg L_a$ and to the short-base linear form for $W \ll L_a$ without separate case analysis.

The boundary value $\delta_0$ is not a free parameter. Applying the law of mass action and quasi-neutrality to both carrier species at an SCR edge gives the exact boundary relation, valid at any bias and injection level,

$$\delta_0^{(n)}(V_j) = \left[\left(\frac{N_D}{2}\right)^2 + n_i^2 e^{V_j/V_T}\right]^{1/2} - \frac{N_D}{2} - p_{n0}\,, \tag{4a}$$

$$\delta_0^{(p)}(V_j) = \left[\left(\frac{N_A}{2}\right)^2 + n_i^2 e^{V_j/V_T}\right]^{1/2} - \frac{N_A}{2} - n_{p0}\,. \tag{4b}$$

Both edges are controlled by the same junction voltage $V_j$: the two boundary populations are locked to each other and cannot be set independently. This relation interpolates continuously between the low-injection Boltzmann form $\delta_0 \simeq p_{n0}(e^{V_j/V_T} - 1)$ and the high-injection form $\delta_0 \simeq n_i e^{V_j/2V_T}$.

***Base field and junction voltage.***—With both carrier species active, the net current $J = J_n + J_p$ fixes the base field self-consistently:

$$E(x) = \frac{J}{\sigma(x)} - \frac{q(D_n - D_p)}{\sigma(x)} \frac{d\delta}{dx} = E_{\text{ohmic}} + E_{\text{Dember}}, \tag{5}$$

$$\sigma(x) = \sigma_0 + q(\mu_n + \mu_p)\,\delta(x), \tag{6}$$

where the conductivity modulation carries the injection dependence. The base field is nonzero throughout the quasi-neutral region, in direct contrast to the classical premise $E = 0$; it separates into an ohmic drift field and a Dember counter-field driven by the unequal electron and hole diffusivities [14-17]. A nonzero base field has two consequences. First, part of the applied bias ($V_a$) is consumed in the bases,

$$V_a = V_j + V_b^{(n)} + V_b^{(p)}, \tag{7}$$

so that $V_j$ and $V_a$ are conceptually and quantitatively distinct. Second, $V_j$ itself is reinterpreted: it is the excess-carrier-induced modification of the built-in barrier, a self-limiting drop confined to the SCR and bounded by the built-in potential $V_0$. The exponential factor $e^{V_j/V_T}$ that drives injection is therefore bounded, and the current remains finite at any applied bias. No external mechanism is needed to keep it so.

***Current components and closure.***—Resolving each species' current into diffusion and drift gives four position-dependent components, three of which (majority diffusion and both drift terms) are absent from the classical treatment. Their position-dependent parts cancel exactly, so $J(x) = J$ at every sectional plane: near the SCR edge the current is primarily carried by the injected minority carriers, and it is progressively transferred to majority-carrier drift toward the contact; there is a continuous minority-to-majority current conversion at constant total current [13]. Current conservation across the SCR (generation–recombination therein neglected, as in the classical theory) then closes the device: the net current equals the sum of the two minority-carrier injection currents evaluated at their own SCR edges,

$$J = J_n(x_p = 0) + J_p(x_n = 0) = G_n\, \delta_0^{(p)}(V_j) + G_p\, \delta_0^{(n)}(V_j), \tag{8}$$

where each injection current includes its diffusion *and* drift parts, and where we define the *injection conversion factors*

$$G_n = \frac{qD_n^{(p)}}{L_a^{(p)}} \coth\frac{W_p}{L_a^{(p)}}, \qquad G_p = \frac{qD_p^{(n)}}{L_a^{(n)}} \coth\frac{W_n}{L_a^{(n)}}, \tag{9}$$

which fold material (diffusivity), geometry (base width through the coth factor), and field coupling (through $L_a$) into a single coefficient converting a boundary excess concentration into a current. Eq. (8) is the *extended diode equation*.

***Correspondence with Shockley's law.***—Under low injection, $\delta_0 \ll N$, the exact boundary excesses reduce to the Boltzmann form, and Eq. (8) becomes $J = (G_n n_{p0} + G_p p_{n0})(e^{V_j/V_T} - 1)$. Under the long-base condition, the coth factor tends to unity, while the ambipolar coefficients reduce to the minority-carrier values, $L_a^{(p)} \to L_n$, $L_a^{(n)} \to L_p$, so that $G_n \to qD_n n_{p0}/L_n$ per unit concentration and likewise for $G_p$. The extended equation thereby reduces systematically to the classical diode equation,

$$J(V_j) = \left(\frac{qD_n n_{p0}}{L_n} + \frac{qD_p p_{n0}}{L_p}\right)\left(e^{V_j/V_T} - 1\right) = J_0\left(e^{V_j/V_T} - 1\right), \tag{10}$$

with the textbook saturation current $J_0$. The Shockley law is thus not replaced but contained: it is the low-injection, long-base limit of Eq. (8), and the regimes in which it fails are precisely those in which its premises fail.

***High injection from first principles.***—Beyond the onset $\delta_0 \approx N$, the boundary excess crosses over to $\delta_0 \simeq n_i e^{V_j/2V_T}$. For an asymmetric p$^+$–n junction the lightly doped side enters high injection first, at a voltage fixed solely by material and doping,

$$V_j^* = 2V_T \ln\frac{N}{2n_i}, \qquad N = \min(N_A, N_D), \tag{11}$$

which defines the onset voltage for high injection, always lies below $V_0$ and delimits the high-injection window $[V_j^*, V_0)$. Within this window the current grows as

$$J \simeq G_p\, \delta_0^{(n)} = \frac{q D_p^{(n)}}{L_a^{(n)}} \coth \frac{W_n}{L_a^{(n)}}\; n_i\, e^{V_j/2V_T}, \tag{12}$$

i.e., with ideality factor n $= 2$, while remaining finite as $V_j \to V_0$. The high-injection roll-off thus follows from first principles, without any empirical series resistance. Fig.2 compares the extended and classical characteristics for a Si p$^+$–n junction over three orders of magnitude in base width: the two laws coincide exactly below $V_j^*$, separate above it, and differ by a factor of $\sim 200$ at $V_j = 0.75$ V, identically for all three widths, because the separation originates in the boundary excess, not the geometry. The base width enters only through $\coth(W/L_a)$: the short-base current is enhanced uniformly by $\approx L_a/W \approx 70$, while the long- and intermediate-base curves nearly overlap.

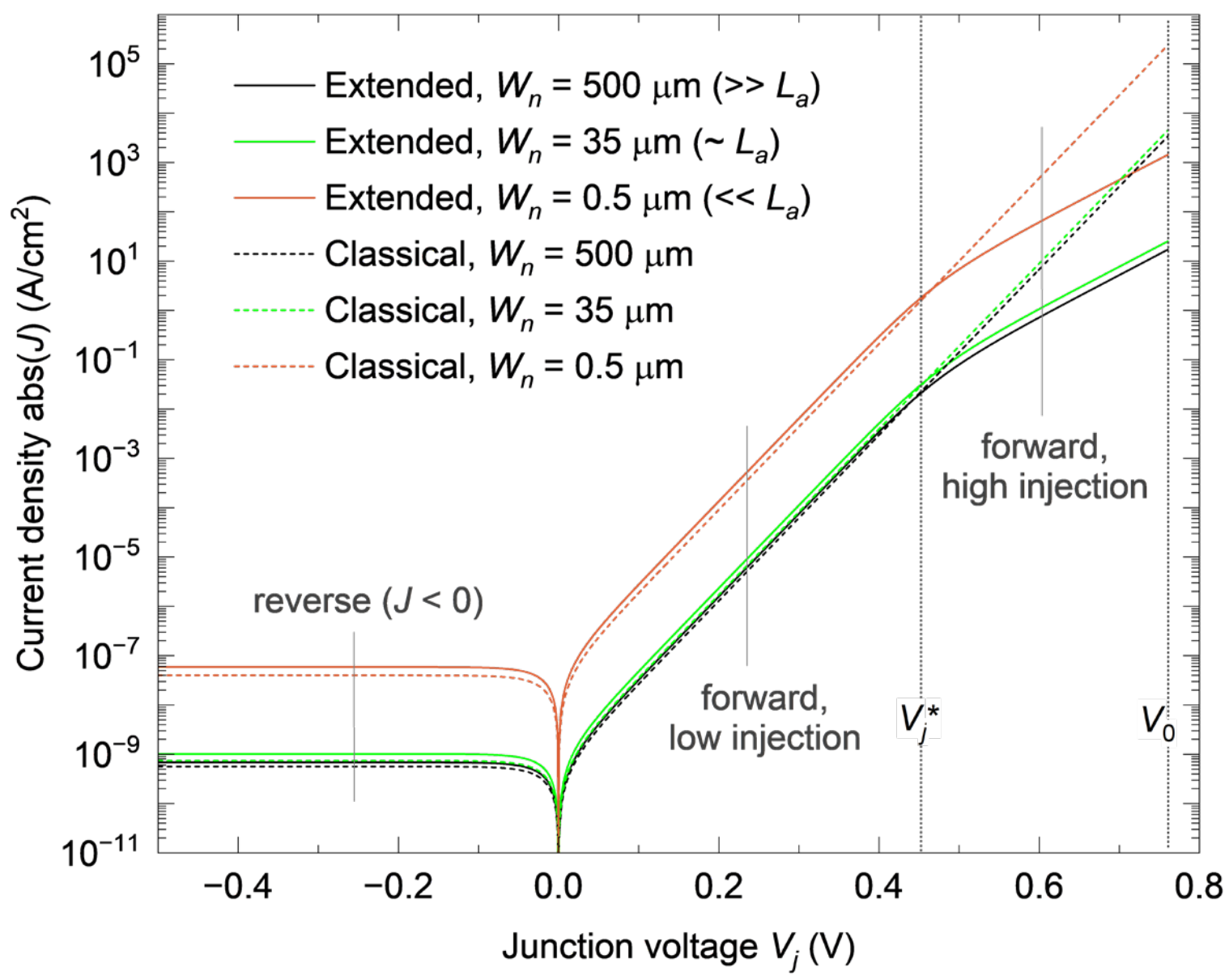

**FIG. 2.** Current density versus junction voltage for a Si p⁺–n junction ($N_A = 5 \times 10^{18}$ cm$^{-3}$, $N_D = 10^{14}$ cm$^{-3}$, $\tau = 1$ μs, 300 K) predicted by the extended diode equation (solid) and the classical diode equation (dashed), for three n-base widths. Dotted and dashed vertical lines mark the onset $V_j^* \approx 0.44$ V and the built-in potential $V_0 \approx 0.76$ V.

***Falsifiable predictions.***—Because every regime is governed by the same boundary relation, the framework makes quantitative predictions that contain no adjustable parameter (Fig.3). (i) *Universal ideality-factor crossover*: Differentiating the boundary relation gives the ideality factor in closed form,

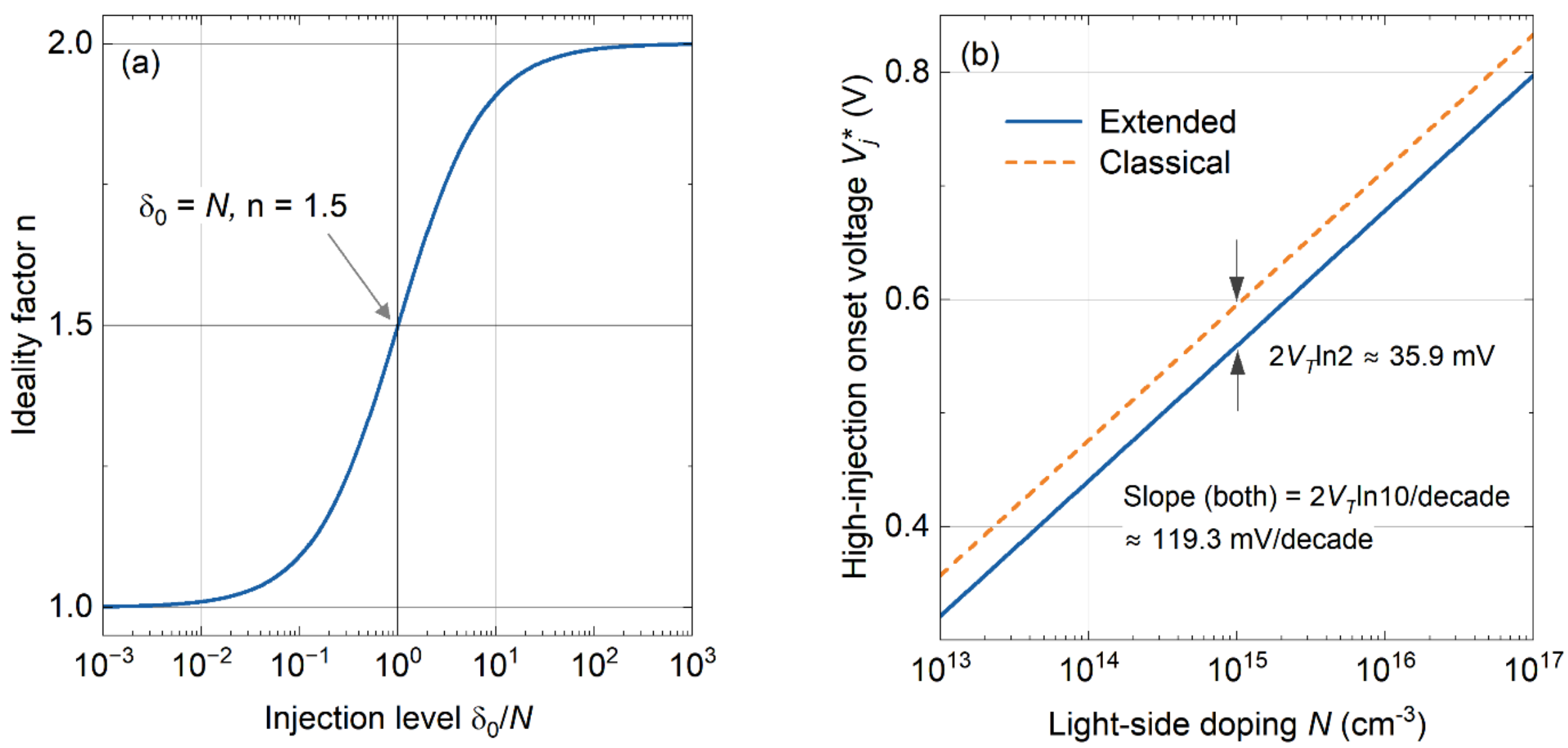

**FIG. 3.** Zero-parameter testable predictions. (a) Universal ideality-factor crossover $\mathrm{n} = (2\delta_0 + N)/(\delta_0 + N)$ versus injection level $\delta_0/N$. (b) Onset voltage $V_j^*$ versus light-side doping (Si, 300 K), with a slope of $2V_T\ln 10 \approx$ 119.3 mV/decade for both classical (orange) and extended (blue) predictions.

$$\mathrm{n} \equiv \left[\frac{d\ln J}{d\left(V_j/V_T\right)}\right]^{-1} = \frac{2\delta_0 + N}{\delta_0 + N}, \tag{13}$$

a universal function of $\delta_0/N$ that evolves from 1 through exactly $3/2$ at $\delta_0 = N$ to 2 (Fig. 3(a)): the $\mathrm{n}(V_j)$ curve measured on any junction should collapse onto this form. (ii) *Doping scaling of the onset voltage*: $V_j^*$ shifts with the light-side doping as $2V_T\ln 10 \approx 119$ mV per decade of $N$ (Fig. 3(b)). (iii) *Geometric factor at high injection*: The short-to-long-base current enhancement $\coth(W/L_a)$ persists unchanged into the $\mathrm{n} = 2$ regime (Fig.2). All three are testable with existing components: a series of $p^+$–n diodes with light-side doping $10^{14}$–$10^{16}$ cm$^{-3}$ and long/short base pairs suffices: measure the forward $J$–$V$, extract $\mathrm{n}(V_j)$, and locate the crossover.

***Relation to earlier work.***—High-level injection phenomena have been studied since the 1950s: Hall and Fletcher analyzed power-rectifier regimes, Herlet gave the classic treatment of p–i–n forward characteristics, and the n = 2 asymptote is recorded in textbooks [6–10,20]. The excess majority carriers themselves are not new to this Letter: our earlier study [22] established that they exist and that their concentration equals that of the excess minority carriers. The accompanying field obtained there, however, is not a complete solution: the net-current closure omitted the minority-carrier drift term, and the treatment was restricted to the long-base, low-injection limit.

The resulting field is therefore approximate even within that regime and meaningless beyond it. Retaining all four current components, including the minority-carrier drift term omitted in that earlier closure, together with an excess-carrier profile valid for arbitrary base length and a boundary relation exact at any injection level, is what makes Eq. (5) an exact closed-form field and what closes the device at Eq. (8). Relative to the classical literature the framework thus differs in kind, not in degree: it supplies (i) a single closed-form law covering all injection levels, bias polarities, and base geometries without case analysis; (ii) an exact self-consistent base field, carried through to the terminal current rather than treated as an internal correction; (iii) a systematic correspondence principle reducing to Shockley's law; and (iv) zero-parameter predictions by which the law stands or falls experimentally. No empirical series resistance enters anywhere: the base drops of Eq. (7) follow from the self-consistent field rather than being fitted.

***Conclusion.***—We have shown that the Shockley diode equation is the low-injection limit of a self-consistent two-carrier drift–diffusion framework in which excess majority carriers, an electrostatically forced population rather than an assumed one, participate in transport, the quasi-neutral base carries a finite field, and the junction voltage is a self-limiting quantity distinct from the applied bias. The framework's device closure, the extended diode equation, contains the classical law exactly, reproduces high-injection roll-off from first principles, and makes three immediately testable predictions. The same base-field and current-component solutions provide the analytical basis for evaluating power dissipation and transit time in junction devices and for improved compact models; the full account is in Ref. [13].